\documentclass[11pt,twoside]{article}
\usepackage{newpasp,epsf}
\def\edcomment#1{\iffalse\marginpar{\raggedright\sl#1\/}\else\relax\fi}
\reversemarginpar

\begin{document}
\title{The Ghosts of Galaxies: Tidal Debris in Clusters}
\author{Michael D. Gregg}
\affil{University of California, Davis and Institute of Geophysics
  and Planetary Physics, Lawrence Livermore National Laboratory }
\author{Michael J. West}
\affil{University of Hawaii, Hilo}

\begin{abstract}
Gravitational interactions in rich clusters can strip material from
the outer parts of galaxies or even completely disrupt entire systems,
giving rise to large scale, low surface brightness ghostly features
stretching across intergalactic space.  The nearby Coma and Centaurus
clusters both have striking examples of galaxy ghosts, in the form of
$100$~kpc-long plumes of intergalactic debris.  By searching HST
archival images, we have found numerous other examples of galaxy
ghosts in rich clusters at low redshift, evidence that galaxy
destruction and recycling are ubiquitous, important in cluster
formation and evolution, and continue to mold clusters at the present
epoch.  Many ghosts appear in X-ray bright clusters, perhaps signaling
a connection with energetic subcluster mergers.

The fate of such material has important ramifications for cluster
evolution.  Our new HST WFPC2 $V \&~I$ images of a portion of the
Centaurus plume reveal that it contains an excess of discrete objects
with $-12 < M_V < -6$, consistent with being globular clusters or
smaller dwarf galaxies.  This tidally liberated material is being
recycled directly into the intracluster population of stars, dwarf
galaxies, globular clusters, and gas, which may have been built
largely from a multitude of similar events over the life of the
cluster.

\end{abstract}

\vspace {-8mm}
\section{Introduction}

Numerous contributions to this symposium have presented exciting
research on intergalactic stellar populations in galaxy clusters.
These populations contribute a substantial fraction of the total light
of a cluster, $20-40\%$ in a cluster as typical as Virgo (see review
by Arnaboldi 2004, this volume).  To understand clusters at even a
basic level, then, it is necessary to account for the presence of this
intergalactic stuff, not only what and how much is there, but how and
when did it get there?  And how does it affect the continued evolution
of the cluster?

A very pretty simulation of cluster formation has been done by
Dubinski (1998; mpg versions available at
www.cita.utoronto.ca/~dubinski/bigcluster.html).  This view of cluster
formation involves not only the gathering together of lots of galaxies
into close quarters, but the agglomeration of a good many into
a single brightest cluster member at the bottom of the global
potential.  This process requires the disruption, destruction, and
recycling of lots of galaxies, giant and dwarf.  The intermediate
products are distended, low surface brightness splashes of stars
across intergalactic space.  Fragile, long filamentary plumes and
extended pools of stars are quickly extruded and dispersed throughout
a cluster, merging into the cD, its extended halo, or the general
intergalactic field.  

\section{Hunting for Galaxy Ghosts}

Cluster formation continues to the present day, so if the Dubinski
simulations are close to the mark, then we should be able to see
examples of these {\em ghosts of galaxies}: the intermediate,
non-equilibrium phases of galaxy disruption and destruction associated
with cluster building.  One spectacular example is the 100~kpc long, low
surface brightness plume in the Coma cluster (Gregg \& West 1998).
Only 10-15~kpc wide, it contains roughly the luminosity of the LMC and
is probably fated to become part of the halo of NGC4874.
While this object seemed unique at first, another plume with nearly
identical properties was soon reported by Calc\'{a}neo--Rold\'{a}n et
al.\ (2000) in the nearby Centaurus cluster.

\begin{figure}[t!]
\plotone{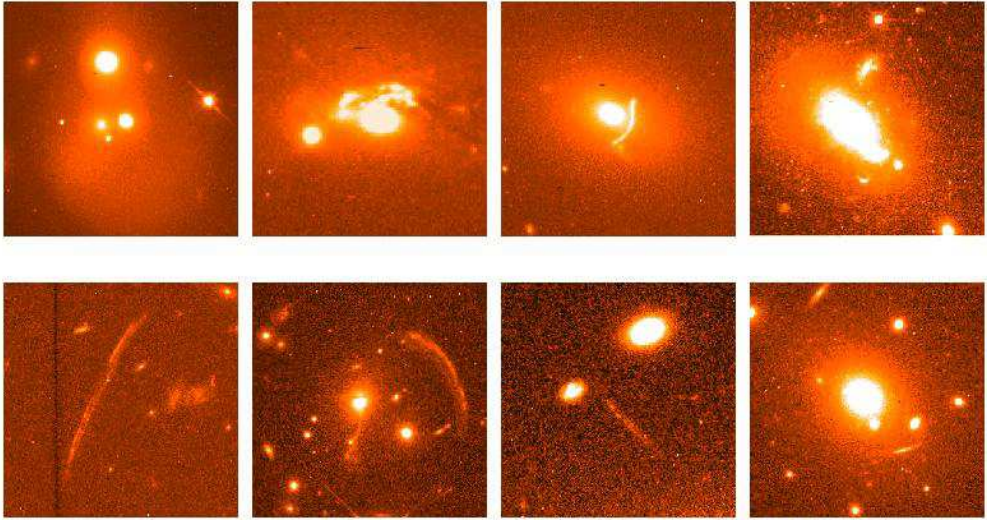}
\caption{\small Examples of tidal debris and/or gravitational arcs
from archival HST/WFPC2 images of clusters with typical $z \approx
0.1$ to 0.3.  The images are 20-30\arcsec\ large.  }
\end{figure}

One example of something in astronomy is an anomaly, but two is a
population, so we combed through the HST archives looking for more
examples of galaxy ghosts.  A number of interesting examples turned
up, especially among X-ray bright clusters; a few are shown in
Figure~1.  An immediately obvious complication is that very thin tidal
plumes can resemble background galaxies gravitationally lensed by the
cluster, and vice versa, especially in ground-based imaging.  Some
clusters exhibit both phenomena; spectroscopic redshifts are needed
for positive identification.  Tidally disrupting galaxies can be any
color: early types contribute old, red stars and globulars to an
intracluster population while gas-rich galaxies may experience a fresh
burst of star formation when pulled apart in a cluster.  Deep
multi-band imaging of these tidal features could provide details of
the stellar and star cluster populations which are being strewn
through intergalactic space.

\section{WFPC2 Observations of the Centaurus Galaxy Ghost}

Analysis of ROSAT images by Churazov et al.\ (1999) shows that the
NGC~4709 and its entourage are violently colliding with the main body
of Centaurus, centered on NGC~4696.  The velocity difference of the
two components is large, $> 1500$~km/s, so perhaps it is no
coincidence that one of the more striking examples of galaxy
disruption is found in Centaurus.  The long, narrow, low surface
brightness plume of material stretches more than 100~kpc across
intergalactic space, arcing past NGC4709 (Figure~2, top, from
Calc\'{a}neo--Rold\'{a}n et al.\ 2000).


To investigate the Centaurus plume in finer detail, we have obtained
WFPC2 $V$ (F606W) and $I$ (F814W) band images of a portion (Figure~2);
the enhanced surface brightness of the plume is clearly visible in the
WFPC2 data (Figure~3).  These images reveal that the plume is not a
completely smooth, low surface brightness feature, but contains dozens
of distinct objects.  Star clusters and dwarf galaxies are known to
form from tidal debris plumes in field galaxies (Duc \& Mirabel 1994;
Duc et al.\ 2000; Whitmore \& Schweizer 1995; Charlton et al.\ 2000),
often accompanied by new star formation.  In Centaurus, a similar
process appears to be at work generating intracluster populations of
stars, globulars, and perhaps even new dwarf galaxies.

\begin{figure}[t!]
\textwidth = 4in
\plotone{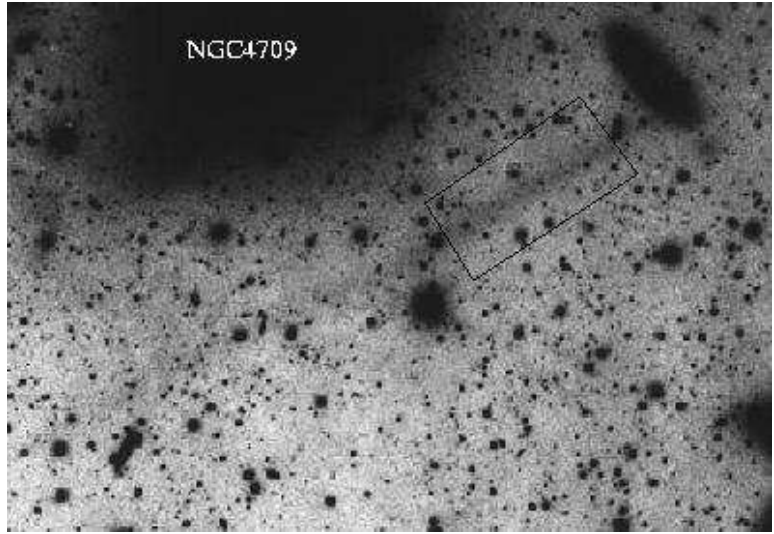}
\caption{\small Discovery image of the Centaurus plume (from
Calc\'{a}neo--Rold\'{a}n et al.\ 2000).  The plume is nearly 8
arcminutes long, over 100~kpc at the distance of Centaurus, extending
from the disky galaxy in the upper right down to the lower left edge
of the field, probably beyond.  The black
rectangle shows the location of our WFPC2 images, chips 2 and 3.}
\end{figure}

\begin{figure}[t!]
\plotone{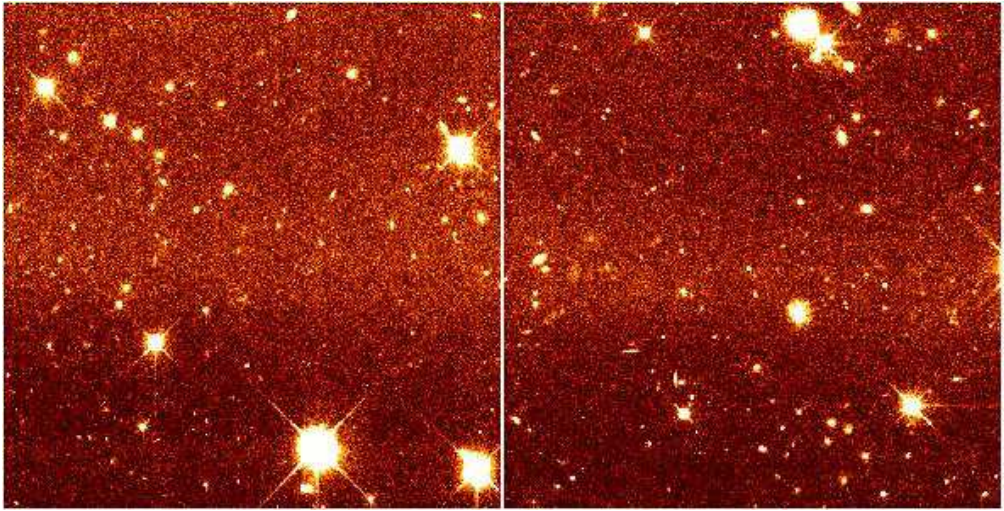}
\caption{\small WFPC2 Chips 2 and 3 of our $V + I$ ($F606W + F814W$)
 observations of the Centaurus tidal debris plume.  Total integration
 time is 4800s in each filter.  The plume is visible as a horizontal stripe of
 enhanced surface brightness.
 A mild gradient imposed by the outer halo of NGC4709 (lying off the
 page to the top) is also present.}
\end{figure}

\vspace{-4mm}
\subsection{Identification of Plume Objects}
\vspace{-1mm}

An object catalog was produced using {\sc sextractor} (Bertin \&
Arnouts 1996) on the summed $V+I$ image.  The numbers of objects as a
function of vertical {\em column} (along the plume in Figure~3) is
statistically consistent with a constant distribution, but the
distribution as a function of horizontal {\em row} (perpendicular to
the plume) is very peaked (Figure~4), differing from
constant with high significance, $> 99.999\%$ in a $\chi^2$ test.  The
median filtered rows and columns of the images (Figure~4, lower panel)
shows that the diffuse low surface brightness light reflects the peak
and shape of the object distribution, confirming the association
between the excess numbers and the plume feature.  A mild gradient
from the outer halo of NGC4709 has been removed from the
median-filtered row plot.

\begin{figure}[t!]
\plottwo{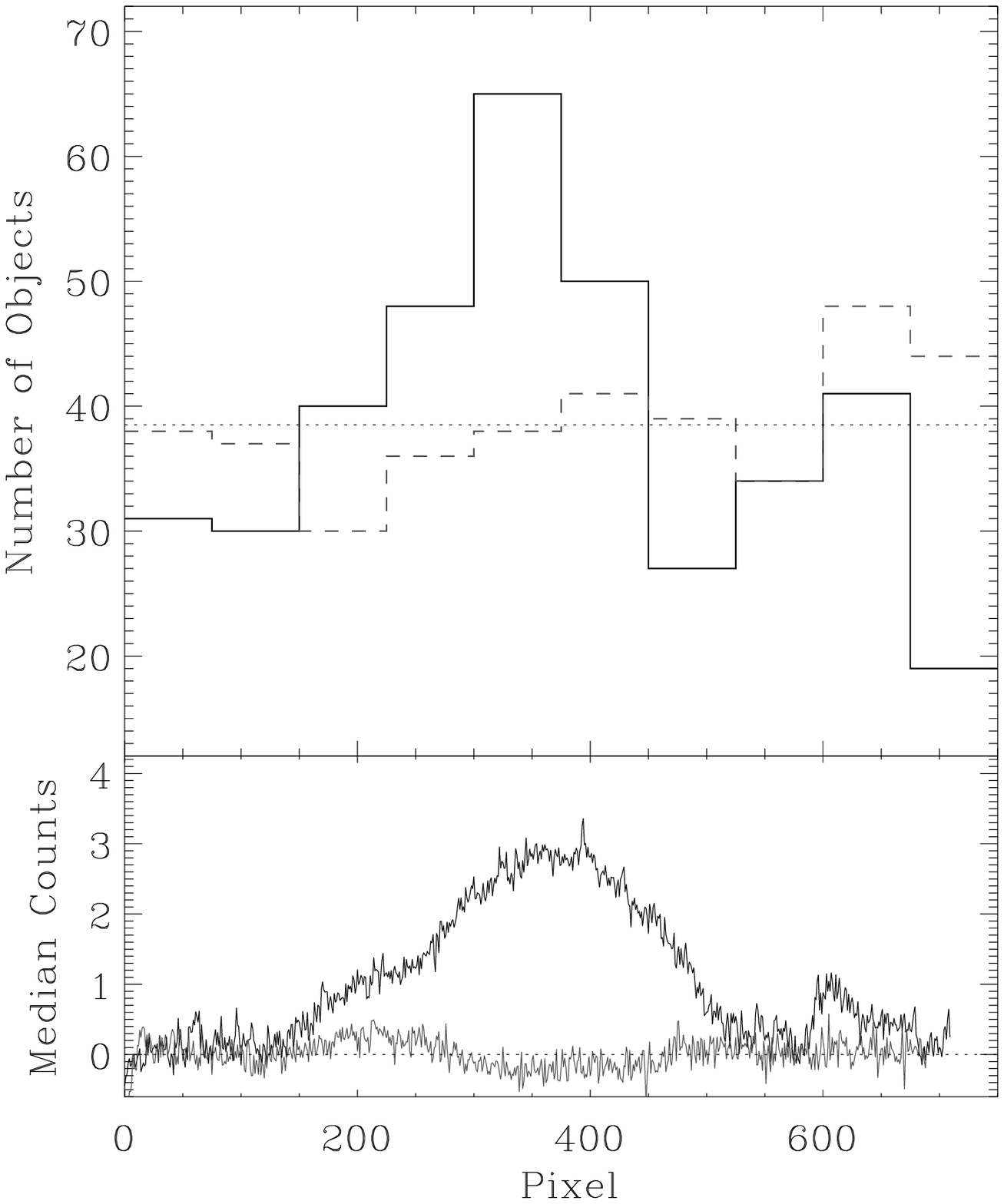}{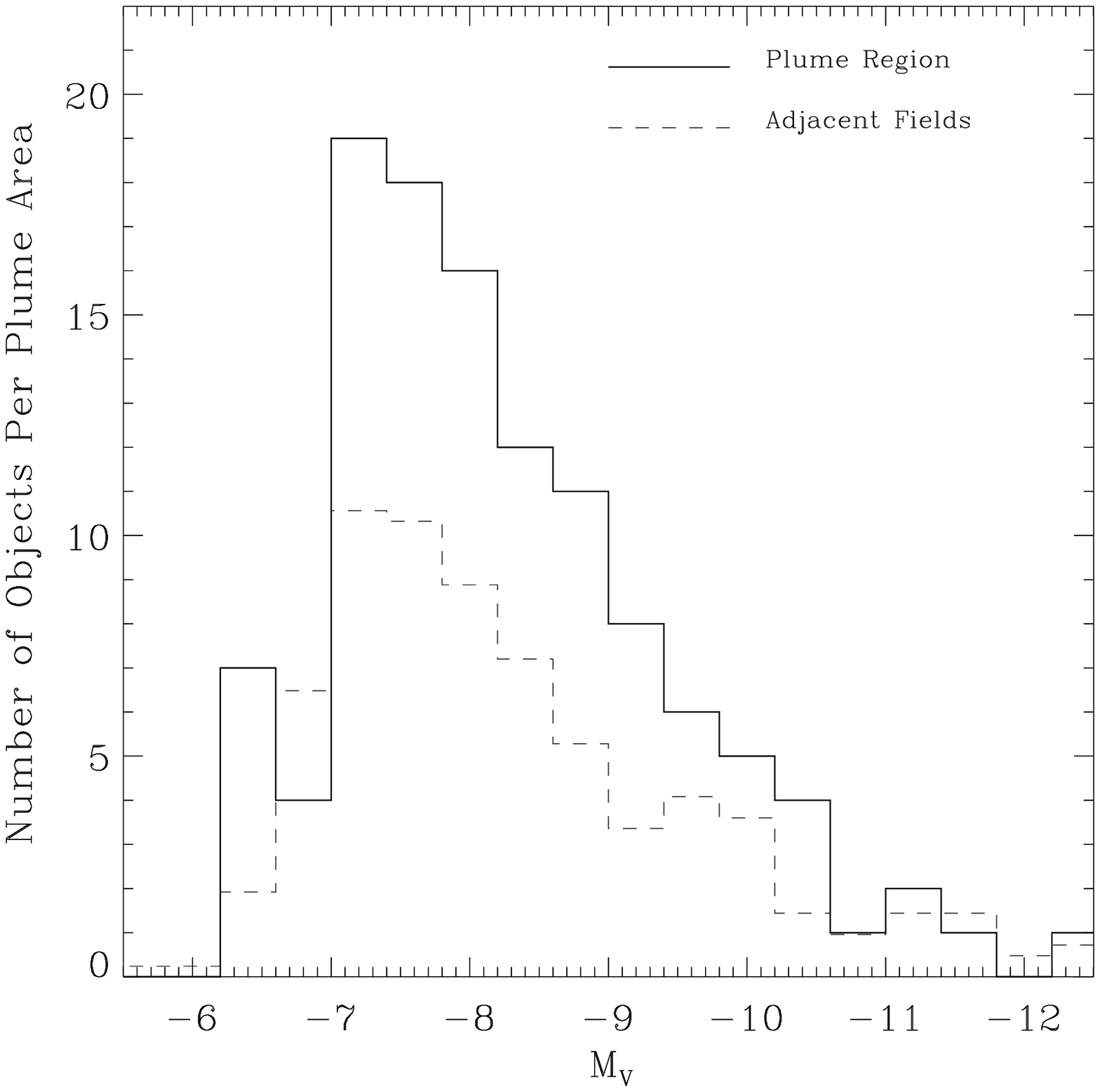}
\caption{\small {\bf Left:} distribution of discrete objects in the
  WFPC2 plume images as a function of row (solid histogram) and column
  number (chips 2 and 3 done separately).  The distribution 
  peaks strongly in the plume, but is roughly constant in the
  orthogonal direction.  The lower plot shows the median of 
  vertical columns (black) and horizontal rows (grey);
  the peak in object numbers occurs in the highest
  surface brightness portion of the plume.  {\bf Right:} Luminosity
  function of objects in and out of the plume; these are in the
  luminosity range of globular clusters and small dwarf galaxies.  }
\end{figure}

\begin{figure}[h!]
\plottwo{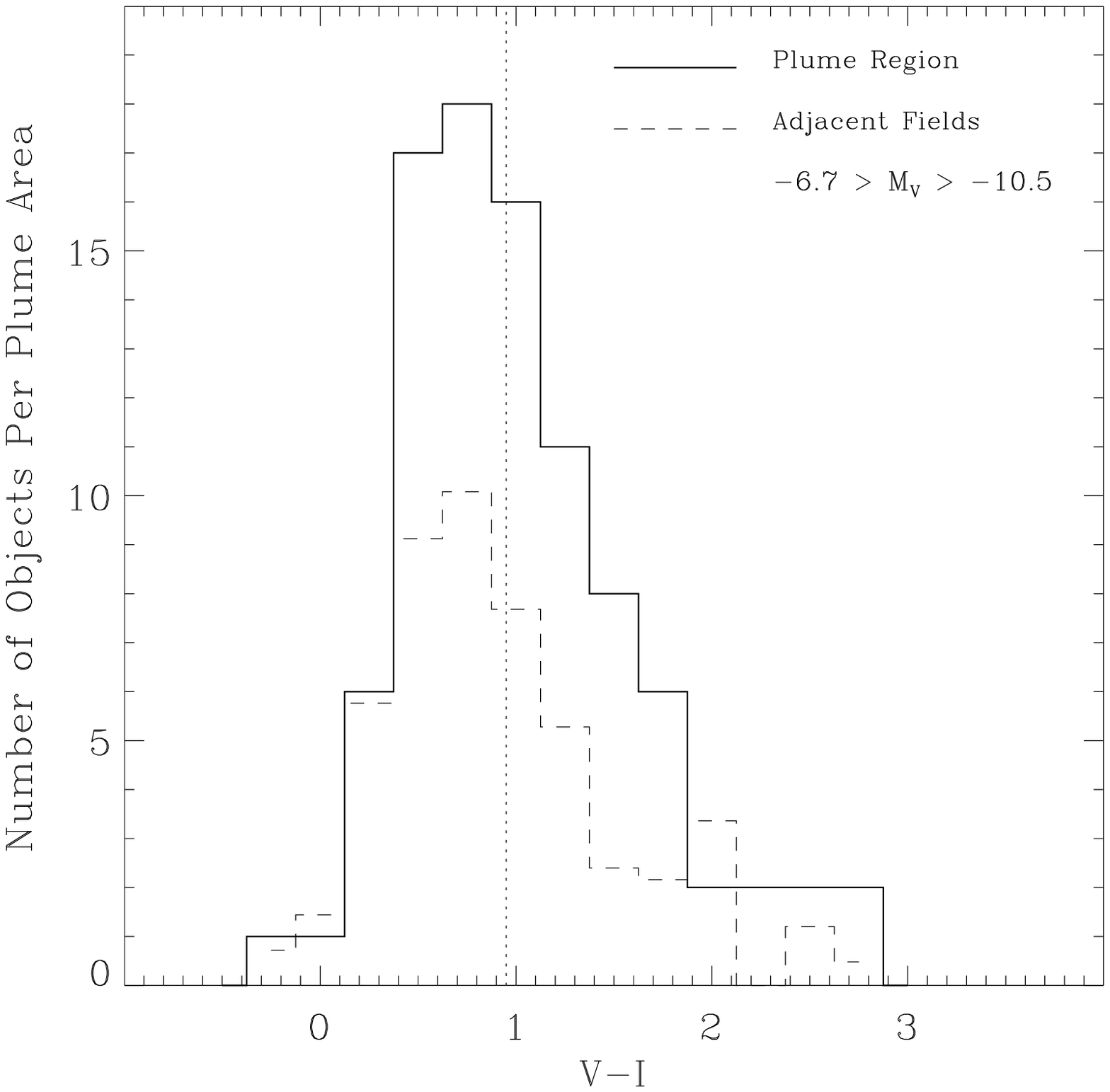}{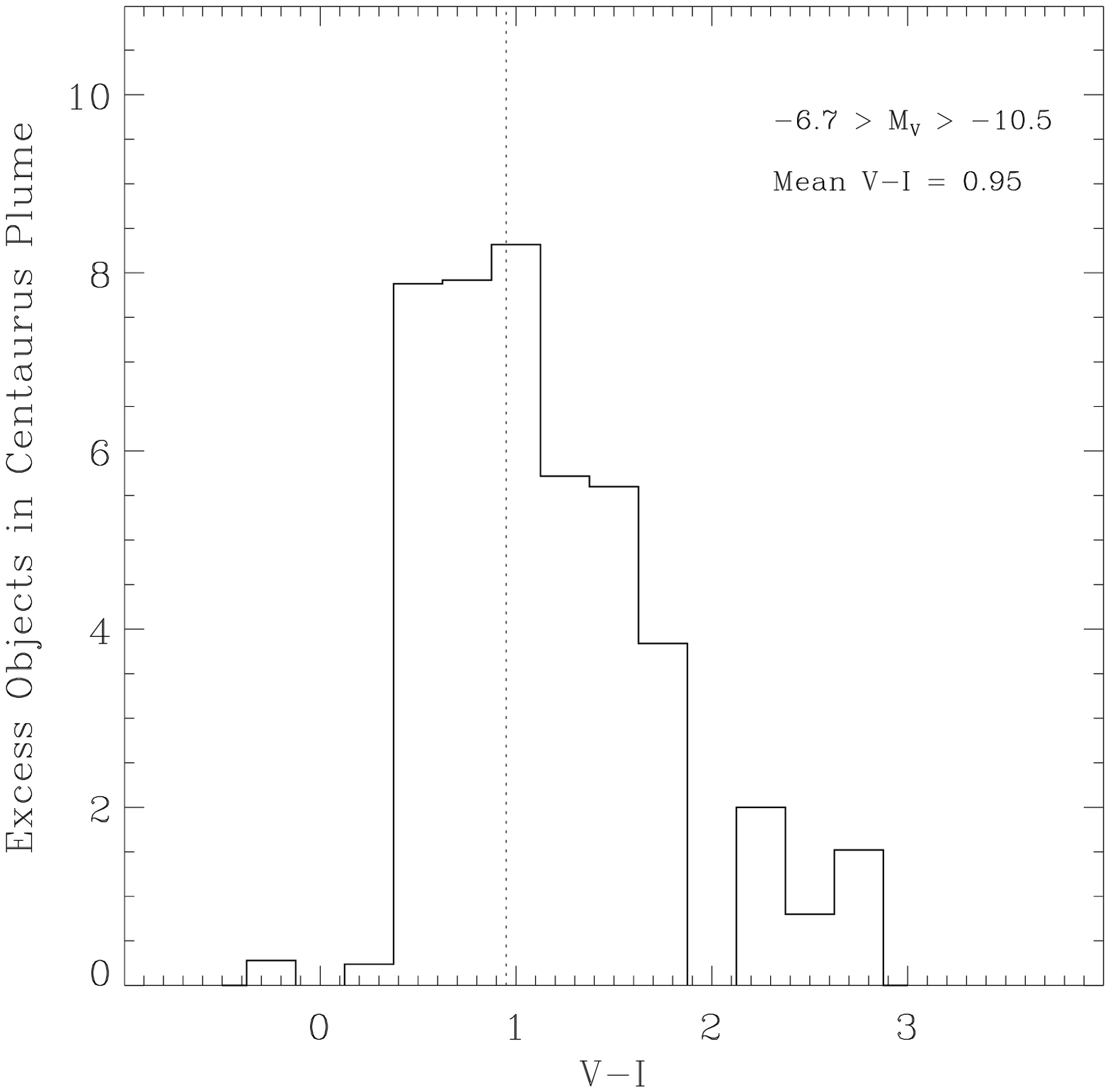}
\caption{\small {\bf Left:} Color-magnitude diagram of plume and
  non-plume objects.  {\bf Right:} Difference histogram showing that
  the plume objects have an average color of $V-I = 0.95$, typical of
  globular clusters in large ellipticals.
}
\end{figure}

\vspace{-1mm}
\subsection{Luminosity Function and Color Magnitude Diagram}
\vspace{-1mm}

The Luminosity function of objects in the plume (Figure~4) compared to
adjacent fields (scaled to an equivalent area) shows that the excess
is mainly in the luminosity range $M_V \approx -7$ to $-9$, but there
are extra objects as bright as $M_V \approx -12$.  (We adopt $M-m =
33.4$, 47~Mpc, for Centaurus.)  These luminosities are in the range of
globular clusters and faint dwarf galaxies; the excess consists of
both extended objects and point sources.  If these objects are bound
and can survive the disruption event, then we are witnessing ongoing
creation of dwarf galaxies from the shards of a larger object and the
injection of old globular star clusters into intergalactic space.

There are 91 objects in the plume, and only $\sim 50$ in the area-scaled
field.  A 'Median-Delta' test shows that the difference of the two is
not random at a significance level of 0.0004.  These numbers predict
that the entire plume (extending far to the east of our WFPC2 field)
contains $\sim 80-100$ objects.

A comparison of the color-magnitude relations for objects in and out
of the plume (Figure~5) shows no discernable difference in color
distributions.  Ordinary globular clusters in these bands have $V-I
\approx 0.95$ (e.g. Kavelaars et al.\ 2000), consistent with the
distribution of plume (and non-plume) object colors.  The plume is in
the outer halo of NGC4709; perhaps many of the non-plume objects are
also globular clusters and faint Centaurus dwarfs.  Most of the
objects are consistent with being point sources, so some could be
individual supergiants, which would imply recent star formation.

\vspace{-3mm}
\section{Discussion and Summary}
\vspace{-2mm}

\begin{itemize}
\item In the Centaurus plume, we are witnessing the present-epoch
creation of intergalactic objects in a rich cluster.  The stripped
material, now floating in the general cluster potential, will
eventually disperse throughout the core of Centaurus, augmenting the
halos of giant ellipticals and the intergalactic populations of stars,
star clusters, and dwarf galaxies.
\item Any interstellar gas and dust associated with these objects is
added to the hot intracluster medium, already bright in the X-ray.
\item The tidal generation of new dwarf galaxies 
can explain the observed association of excess faint galaxies with
dynamically turbulent clusters (e.g.\ Lopez-Cruz et al.\ 1997). 
\item Disruption of infalling galaxies and the liberation of
tidal debris is an important driver of the evolution of member
galaxies the cluster as a whole.
\item Objects such as the tidal plume in Centaurus provide a glimpse
of a more chaotic era when today's rich galaxy clusters were beginning
to form.
\end{itemize}

\acknowledgements {
This work is based on observations made with the
Hubble Space Telescope; support was provided by NASA
through grant GO-8644 from the Space Telescope Science
Institute, operated by AURA, Inc., under NASA contract
NAS5-26555.  Part of this work was performed under the auspices of the
U.S. Department of Energy by University of California Lawrence
Livermore National Laboratory under contract No.~W-7405-Eng-48. }

\end{document}